# Identifying lineage effects when controlling for population structure improves power in bacterial association studies

Sarah G Earle[1*], Chieh-Hsi Wu[1*], Jane Charlesworth[1*], Nicole Stoesser[1], N Claire Gordon[1], Timothy M Walker[1], Chris C A Spencer[2], Zamin Iqbal[2], David A Clifton[3], Katie L Hopkins[4], Neil Woodford[4], E Grace Smith[5], Nazir Ismail[6], Martin J Llewelyn[7], Tim E Peto[1], Derrick W Crook[1], Gil McVean[2], A Sarah Walker[1], Daniel J Wilson[1,2]

Addresses:
[1] Nuffield Department of Medicine, University of Oxford, John Radcliffe Hospital, Oxford, OX3 9DU, United Kingdom
[2] Wellcome Trust Centre for Human Genetics, University of Oxford, Roosevelt Drive, Oxford, OX3 7BN, United Kingdom
[3] Institute of Biomedical Engineering, Department of Engineering Science, University of Oxford, Oxford, United Kingdom
[4] Antimicrobial Resistance and Healthcare Associated Infections Reference Unit, Public Health England, London, NW9 5EQ, United Kingdom
[5] Public Health England, West Midlands Public Health Laboratory, Heartlands Hospital, Birmingham, B9 5SS, United Kingdom
[6] Centre for Tuberculosis, National Institute for Communicable Diseases, Johannesburg, South Africa
[7] Department of Infectious Diseases and Microbiology, Royal Sussex County Hospital, Brighton, United Kingdom

[*] These authors contributed equally
Corresponding author D.J.W. Email: daniel.wilson@ndm.ox.ac.uk

## Abstract
Bacteria pose unique challenges for genome-wide association studies (GWAS) because of strong structuring into distinct strains and substantial linkage disequilibrium across the genome. While methods developed for human studies can correct for strain structure, this risks considerable loss-of-power because genetic differences between strains often contribute substantial phenotypic variability. Here we propose a new method that captures lineage-level associations even when locus-specific associations cannot be fine-mapped. We demonstrate its ability to detect genes and genetic variants underlying resistance to 17 antimicrobials in 3144 isolates from four taxonomically diverse clonal and recombining bacteria: *Mycobacterium tuberculosis*, *Staphylococcus aureus*, *Escherichia coli* and *Klebsiella pneumoniae*. Strong selection, recombination and penetrance confer high power to recover known antimicrobial resistance mechanisms, and reveal a candidate association between the outer membrane porin *nmpC* and cefazolin resistance in *E. coli*. Hence our method pinpoints locus-specific effects where possible, and boosts power by detecting lineage-level differences when fine-mapping is intractable.

## Introduction
Mapping genetic variants underlying bacterial phenotypic variability is of great interest owing to the fundamental role of bacteria ecologically, industrially and in the global burden of disease[1-6]. Hospital-associated infections including *Staphylococcus aureus, Escherichia coli* and *Klebsiella pneumoniae* represent a serious threat to the safe provision of healthcare[7-9], while the *Mycobacterium tuberculosis* pandemic remains a major global health challenge[10]. Treatment options continue to be eroded by the spread of antimicrobial resistance, with some strains resistant even to antimicrobials of last resort[11].

GWAS offers new opportunities to map bacterial phenotypes through inexpensive sequencing of entire genomes enabling direct analysis of causal loci, and functional validation via well-developed molecular approaches[12-27]. However, bacterial populations typically exhibit genome-wide linkage disequilibrium (LD) and strong structuring into geographically-widespread genetic lineages or strains that are likely maintained by natural selection[28,29]. Approaches to controlling for this population structure have allowed for systematic phenotypic differences based on cluster membership[20,21] or, in clonal species, phylogenetic history[18,24-26]. However, these and other approaches common in human GWAS[30-32] risk masking causal variants because differences between strains account for large proportions of both phenotypic and genetic variability.

Here we describe a new approach for controlling bacterial population structure that boosts power by recovering signals of lineage-level associations when associations cannot be pinpointed to individual loci because of strong population structure, strong LD and lack of homoplasy. We base our approach on linear mixed models (LMMs), which can control for close relatedness within samples by capturing the fine structure of populations more faithfully than other approaches[33-36], and enjoy greater applicability than phylogenetic methods since recombination is evident in most bacteria[37,38]. Our approach offers biological insights into strain-level differences and identifies groups of loci that are collectively significant even when individually insignificant, without sacrificing power to detect locus-specific associations.

## Results

**Controlling for population structure risks widespread loss of power in bacteria**
Controlling for population structure aims to avoid spurious associations arising from (i) LD with genuine causal variants that are population-stratified, (ii) uncontrolled environmental variables that are population-stratified, and (iii) population-stratified differences in sampling[31,32]. In the four species we investigated, we observed genome-wide LD and strong population structure, with the first 10 principal components (PCs[39]) explaining 70-93% of genetic variation, compared to 27% in human chromosome 1 (**Supplementary Fig. 1**). Controlling artefacts arising from population structure therefore risks loss of power to detect genuine associations in this large proportion of population-stratified loci.

For example, we investigated associations between fusidic acid resistance and the presence or absence of short 31bp haplotypes or *kmers* in *S. aureus* (**Supplementary Fig. 2**). The kmer approach aims to capture resistance encoded by substitutions in the core genome, the presence of mobile accessory genes, or both[18]. Kmers linked to the presence of *fusC*, a mobile-element-associated resistance-conferring gene whose product prevents fusidic acid interacting with its target EF-G[40-43], showed the strongest genome-wide association by $\chi^2$ test ($p = 10^{-122}$).

However, *fusC*-encoded resistance was observed exclusively within strains ST-1 and ST-8. Thus, controlling for population structure using LMM[44,45] reduced significance to $p = 10^{-39}$, below other loci (**Fig. 1a**, **Supplementary Fig. 3**). Kmers capturing resistance-conferring substitutions in *fusA*, which encodes EF-G, were propelled to greater significance because these low-frequency variants were unstratified, and LMM improves power in the presence of polygenic effects[46] ($p = 10^{-11}$ by $\chi^2$ test, $p = 10^{-157}$ by LMM). However, *fusA* variants explain only half as much resistance as *fusC* overall.

While kmers linked to *fusC* did not suffer outright loss of significance, as penetrance was very high, simulations show that for phenotypes with modest effect sizes (e.g. odds ratios of 3), controlling for population structure risks loss of genome-wide significance at 59%, 75%, 99% and 99% of high-frequency causal variants in *M. tuberculosis* ($n = 1573$), *S. aureus* ($n = 992$), *E. coli* ($n = 241$) and

*K. pneumoniae* (*n* = 176) simulations respectively, with power loss greatest when sample size is low and the number of variants is high (**Fig. 2a, Supplementary Fig. 4a**).

**Detecting lineage- versus locus-specific associations**

Methods to limit loss of power such as leave-one-chromosome-out[46,47] are impractical in bacteria, which typically have one chromosome. Instead we developed a method to recover information discarded when controlling for population structure. In cases where population stratification reduces power to detect locus-specific associations, our method infers lineage-specific associations, similar to a phylogenetic regression[48-50], without sacrificing power to detect locus-specific associations when able to do so.

We observed that leading PCs tend to correspond to major lineages in bacterial genealogies (or *clonal frames*[51-53]) despite substantial differences in recombination rates (**Fig. 1b, Supplementary Fig. 5**), reflecting an underlying relationship between genealogical history and PC analysis[54]. PCs are commonly used to control for population structure by including leading PCs as fixed effects in a regression[39]. Therefore regression coefficients estimated for PCs could be interpreted as capturing lineage-level phenotypic differences, and each PC tested for an effect on the phenotype. Since PCs are guaranteed to be uncorrelated, defining lineages in terms of PCs, rather than as phylogenetic branches or genetic clusters, minimizes loss-of-power to detect lineage-level associations caused by correlations between lineages.

To identify lineage effects we exploited a connection between PCs and LMMs. In an LMM, every locus is included as a random effect in a regression. This is equivalent to including every PC in the regression as a random effect[55]. Therefore we decomposed the random effects estimated by the LMM to obtain coefficients and standard errors for every PC (see Methods). We then employed a Wald test[56] to assess the significance of the association between each lineage and the phenotype.

Our method, implemented in the R package *bugwas*, revealed strong signals of association between fusidic acid resistance and lineages including PC-6 and PC-9 ($p = 10^{-70}$), comparable in significance to the low-frequency variants at *fusA* (**Fig. 1c, Supplementary Fig. 6**). We next reassessed locus-specific effects by assigning variants to lineages according to the PC to which they were most correlated, and comparing the significance of variants within lineages. This showed that *fusC* and variants in LD with *fusC* accounted for the strongest signals within PC-6 and PC-9 ($p = 10^{-34}$ and $10^{-45}$ respectively) (**Fig. 1d**), with the strongest locus-specific associations localized to a 20kb region containing the staphylococcal cassette chromosome (SCC), the most significant hit mapping to the gene adjacent to *fusC*. Thus identifying loci contributing to the most significant lineages provides an alternative to prioritizing variants for follow-up based solely on locus-specific significance.

In simulations, our method was able to recover signals of lineage-level associations in cases where significance at individual loci was lost by controlling for population structure, increasing power 2.5- (*M. tuberculosis*) to 22.0-fold (*E. coli*) (**Fig. 2a, Supplementary Fig. 4a**). LMM reduced the number of falsely detected SNPs by 30- (*K. pneumoniae*) to 3600-fold (*S. aureus*). However, fine-mapping of causal variants to specific chromosomal regions frequently suffered from genome-wide LD because LD is not generally organized into physically linked blocks along the chromosome (**Fig. 2b, Supplementary Fig. 4b**), underlining the importance of recovering power by interpreting lineage effects.

We noted a trade-off to interpreting lineage effects because they are susceptible to confounding with population-stratified differences in environment or sampling (**Supplementary Fig. 7**). Therefore, non-random associations between lineages and uncontrolled variables that influence phenotype risks false detection of lineage-level differences.

## Multi-drug resistance and multisite resistance mechanisms influence signals of association

Confronted with strong population structure and genome-wide LD in bacteria, we wished to test empirically the ability of GWAS to pinpoint genuine causal variants more generally. Therefore we conducted 26 GWAS for resistance to 17 antimicrobials in 3144 isolates across the major pathogens *M. tuberculosis*[57], *S. aureus*[58], *E. coli* and *K. pneumoniae*[59] (**Supplementary Fig. 8**).

We supplemented the kmer approach by surveying variation in SNPs and gene presence or absence. We imputed missing SNP calls by reconstructing the clonal frame[51,53] followed by ancestral state reconstruction[61], an approach that generally outperformed imputation using Beagle[60] (**Supplementary Table 1**).

Correlated phenotypes caused by the presence of multi-drug resistant isolates led to significant results in unexpected loci or regions in some analyses. A combination of first-line drug regimens contributes to multi-drug resistance co-occurrence in *M. tuberculosis*, which led to spurious associations as the top hit before controlling for population structure between ethambutol and pyrazinamide resistance and SNPs in rifampicin resistance-conferring *rpoB*. Even after controlling for population structure, these associations remained genome-wide significant at $p = 10^{-45}$ and $p = 10^{-54}$.

Antimicrobial resistance has arisen over 20 times per drug in the *M. tuberculosis* tree, through frequent convergent evolution (**Supplementary Fig. 4c**, **Supplementary Fig. 8**). Within a single gene, such as *rpoB*, there are multiple targets for selection. Both SNP and kmer-based approaches correctly identified variants in known resistance-causing codons, but greater significance was attained in the latter since the targets for selection were typically within 31bp (**Supplementary Fig. 9a**). In these cases, absence of the wild type allele was found to confer resistance, with power gained by pooling over the alternative mutant alleles.

## Strong selection and recombination assist fine-mapping of antibiotic resistance determinants

For each drug and species, we evaluated whether the most significant hit identified by GWAS matched a known causal variant[57-59] (**Supplementary Table 2**). By this measure, the performance of GWAS across species was very good, identifying genuine causal loci or regions in physical linkage with those loci for antimicrobial resistance in 25/26 cases for the SNP and gene approach and the kmer approach after controlling for population structure (**Table 1, Supplementary Table 3**). Particularly for accessory genes such as β-lactamases, mobile element-associated regions of LD were often detected along with the causal locus (**Supplementary Figure 9b**).

Genuine resistance-conferring variants were detected in all but one study, demonstrating that the high accuracy attained in predicting antimicrobial resistance phenotypes from genotypes known from the literature[58,62] is mirrored by good power to map the genotypes that confer antimicrobial resistance phenotypes using GWAS. However, these results also reflect extraordinary selection pressures exerted by antimicrobials. High homoplasy at resistance-conferring loci caused by repeat mutation and recombination break down LD, assisting mapping (**Fig. 2c, Supplementary Fig. 4c**).

For one drug, cefazolin, in *E. coli*, we identified variation in the presence of an unexpected gene as the most strongly associated with resistance, *nmpC* ($p = 10^{-12.4}$). This gene encodes an outer membrane porin over-represented in susceptible individuals. Permeability in the *Salmonella typhimurium* homolog mediates resistance to other cephalosporin beta-lactams[63], making this a strong candidate for a novel resistance-conferring mechanism discovered in *E. coli*.

## Conclusions

Population structure presents the greatest challenge for GWAS in bacteria, because of the inherent trade-off between power to detect genuine associations of population-stratified variants and robustness to unmeasured, population-stratified confounders. By introducing a test for lineage-specific associations, we allow these signals to be recovered even in the absence of homoplasy, while acknowledging the increased risk of confounding. Detecting lineage effects is valuable because characterizing phenotypic variability in terms of strain-level differences is helpful for biological understanding and it permits the prediction of traits, including clinically actionable phenotypes, from strain designation.

Identifying loci that contribute to the most significant lineage-level associations offers flexibility in the interpretation of bacterial GWAS, where it will often be difficult to pinpoint significance to individual locus effects, and where LD can make the fine-mapping of causal loci a genome-wide problem. Loci can be prioritized for follow-up by identifying groups of lineage-associated variants that collectively show a strong signal of phenotypic association, but which cannot be distinguished statistically. This strategy provides an alternative to prioritizing variants based solely on locus-specific significance, but it carries risks, because lineage-associated effects are more susceptible to confounding with population-stratified differences in environment or sampling. This trade-off between power and robustness underlines the importance of functional validation for bacterial GWAS going forward.

## Methods

**Linear mixed model.** In the linear mixed model[44-47] (LMM), the phenotype is modelled as depending on the fixed effects of covariates including an intercept, the "foreground" fixed effect of the locus whose individual contribution is to be tested, the "background" random effects of all the loci whose cumulative contribution to phenotypic variability we will decompose into lineage-level effects, and the random effect of the environment:

$$phenotype = covariates + foreground\ locus + background\ loci + environment$$

Formally,

$$y_i = W_{i1}\alpha_1 + \ldots + W_{ic}\alpha_c + X_{il}\beta_l + X_{i1}\gamma_1 + \ldots + X_{iL}\gamma_L + \varepsilon_i,$$

where there are $n$ individuals, $c$ covariates, $L$ loci, $l$ is the foreground locus, $y_i$ is the phenotype in individual $i$, $W_{ij}$ is covariate $j$ in individual $i$, $\alpha_j$ is the effect of covariate $j$, $X_{ij}$ is the genotype of locus $j$ in individual $i$, $\beta_l$ is the foreground effect of locus $l$, $\gamma_j$ is the background effect of locus $j$ and $\varepsilon_i$ is the effect of the environment (or error) on individual $i$. Biallelic genotypes are numerically encoded as $-f_j$ (common allele) or $1-f_j$ (rare allele), where $f_j$ is the frequency of the rare allele at locus $j$. This convention ensures the mean value of $X_{ij}$ over individuals $i$ is zero for any locus $j$. Since triallelic and tetrallelic loci are rare, we use only biallelic loci to model background effects. When the foreground locus is triallelic ($K = 3$) or tetrallelic ($K = 4$), the genotype in individual $i$ is encoded as a vector indicating the presence (1) or absence (0) of the first ($K-1$) alleles and $\beta_l$ becomes a vector of length ($K-1$).

Treating the background effects of the loci as random effects means the precise values of the coefficients $\gamma_j$ are averaged over. The $\gamma_j$s are assumed to follow independent normal distributions with common mean 0 and variance $\lambda\tau^{-1}$ to be estimated. Since most loci are expected to have little or no effect on a particular phenotype, this tends to constrain the magnitude of the background effect sizes to be small. The environmental effects are also treated as random effects assumed to follow independent normal distributions with mean 0 and variance $\tau^{-1}$. The model can be rewritten in matrix form as

$$\mathbf{y} = \mathbf{W}\boldsymbol{\alpha} + \mathbf{X}_{\cdot l}\beta_l + \mathbf{u} + \boldsymbol{\varepsilon}$$

with

$$\mathbf{u} = \mathbf{X}_{\cdot 1}\gamma_1 + \ldots + \mathbf{X}_{\cdot L}\gamma_L$$

$$\mathbf{u} \sim \text{MVN}_n(0, \lambda\, \tau^{-1}\, \mathbf{K})$$
$$\boldsymbol{\varepsilon} \sim \text{MVN}_n(0, \tau^{-1}\, \mathbf{I}_n)$$

where $\mathbf{u}$ represents the cumulative background effects of the loci, MVN denotes the multivariate normal distribution, $\mathbf{I}_n$ is an $n \times n$ identity matrix and $\mathbf{K}$ is an $n \times n$ relatedness matrix defined as $\mathbf{K} = \mathbf{X}\,\mathbf{X}'$, which captures the genetic covariance between individuals.

**Testing for locus effects.** To assess the significance of the effect of an individual locus $l$ on the phenotype, controlling for population structure and background genetic effects, the parameters of the linear mixed model $\alpha_1\ldots\alpha_c$, $\beta_l$, $\lambda$ and $\tau$ were estimated by maximum likelihood and a likelihood ratio test with $(K-1)$ degrees of freedom was performed against the null hypothesis that $\beta_l = 0$ using the software GEMMA[45].

**Testing for lineage effects.** Since controlling for population structure drastically reduces power at population-stratified variants, and since a large proportion of variants are typically population-stratified in bacteria, we recovered information from the LMM regarding lineage-level differences in phenotype.

We defined lineages using principal components (PCs) because we observed that PCs tend to trace paths through the clonal frame genealogy corresponding to recognizable lineages and because PCs are mutually uncorrelated, minimizing loss-of-power to detect differences between lineages due to correlations. PCs were computed based on biallelic SNPs using the R function prcomp(), producing an $L$ by $n$ loading matrix $\mathbf{D}$ and an $n$ by $n$ score matrix $\mathbf{T}$ where $\mathbf{T} = \mathbf{X}\,\mathbf{D}$. $D_{ij}$ records the contribution of biallelic SNP $i$ to the definition of PC $j$ while $T_{ij}$ represents the projection of individual $i$ on to PC $j$.

Point estimates and standard errors for the background locus effects are usually overlooked because the assumed normal distribution with common mean 0 and variance $\lambda\tau^{-1}$ tends to cause them to be small in magnitude and not significantly different from zero. However, cumulatively the background locus effects can capture systematic phenotypic differences between lineages. Therefore we recovered the post-data distribution (equivalent to an empirical Bayes posterior distribution) of the background locus random effects, $\boldsymbol{\gamma}$, from the LMM, and reinterpreted it in terms of lineage-level differences in phenotype.

Empirically, we found that the post-data distribution of the background random effects was generally insensitive to the identity of the foreground locus and comparable under the null hypothesis ($\beta_l = 0$). Therefore, we calculated the mean and variance-covariance matrix of the multivariate normal post-data distribution of $\boldsymbol{\gamma}$ in the LMM null model. These are equivalent to those of a ridge regression[64], and were computed as
$$\boldsymbol{\mu} = (\mathbf{X}'\mathbf{X} + 1/\lambda\, \mathbf{I}_L)^{-1}\, \mathbf{X}'\mathbf{y} \text{ and } \boldsymbol{\Sigma} = (\tau\,\mathbf{X}'\mathbf{X} + 1/\lambda\, \mathbf{I}_L)^{-1}$$
respectively. Both $\lambda$ and $\tau$ were estimated by GEMMA under the LMM null model.

Using the inverse transformation of the biallelic variants from PCA, $\mathbf{X} = \mathbf{T}\,\mathbf{D}^{-1}$, the background random effects can be rewritten in terms of the contribution of the $n$ PCs
$$\mathbf{u} = \mathbf{X}_{\cdot 1}\gamma_1 + \ldots + \mathbf{X}_{\cdot L}\gamma_L$$
$$= \mathbf{X}\boldsymbol{\gamma} = \mathbf{T}\,\mathbf{D}^{-1}\boldsymbol{\gamma} = \mathbf{T}\,\mathbf{g}$$
$$= \mathbf{T}_{\cdot 1}g_1 + \ldots + \mathbf{T}_{\cdot n}g_n$$
where $\mathbf{g} = \mathbf{D}^{-1}\boldsymbol{\gamma}$, $g_j$ being the background effect of PC $j$ on the phenotype. We therefore computed the mean and variance of the post-data distribution of $\mathbf{g}$ as $\mathbf{m} = \mathbf{D}^{-1}\boldsymbol{\mu}$ and $\mathbf{S} = \mathbf{D}^{-1}\boldsymbol{\Sigma}\mathbf{D}$ respectively using the affine transformation for a multivariate normal distribution. To test the null hypothesis of no background effect of PC $j$ (i.e. $g_j = 0$) we employed a Wald test with test statistic $w_j = m_j^2/S_{jj}$, which we compared against a chi-squared distribution with one degree of freedom to obtain a $p$-value.

While we identified and tested for lineage effects in the LMM setting, lineage effects could also be identified and tested for by interpreting the coefficients of leading PCs or genetic cluster membership included as fixed effects in a regression, both of which represent alternative methods for controlling for population structure.

**Identifying non genome-wide PCs.** Some PCs capture variation localized to particular areas of the genome. We identified non genome-wide PCs by testing for spatial heterogeneity of the loading matrix **W** for biallelic SNPs across the genome. SNPs were grouped into 20 contiguous bins (indexed by $j$) of nearly equal sizes $N_j$ and the mean $O_{ij}$ and variance $V_{ij}$ in the absolute value of the SNP loadings for PC $i$ in bin $j$ were calculated, along with the mean absolute value $E_i$ of the SNP loadings for PC $i$ across all SNPs. The null hypothesis of no heterogeneity was assessed by comparing the test statistic $\chi_i^2 = \Sigma_j (O_{ij} - E_i)^2/(V_{ij}/N_j)$ to a chi-squared distribution with degrees of freedom equal to the number of bins minus one to obtain a *p*-value.

**Antimicrobial resistance testing, genome sequencing and SNP calling.** We investigated 241 *E. coli* and 176 *K. pneumoniae* UK clinical isolates newly reported here, together with 992 *S. aureus* and 1735 *M. tuberculosis* isolates reported previously[57,58]. All isolates were tested for resistance to multiple antimicrobials based on routine clinical laboratory protocols and DNA was extracted and sequenced on Illumina platforms as previously described[57-59]. We called SNPs using standard methods[65,66], employing Stampy[67] to map reads to reference strains CFT073 (genbank accession AE014075.1), MGH 78578 (CP000647.1), H37Rv (NC_000962.2) and MRSA252 (BX571856.1) for *Escherichia coli, Klebsiella pneumoniae, Mycobacterium tuberculosis and Staphylococcus aureus* respectively. All genomes were deposited in NCBI and EBI short read archives under project accession numbers PRJNA306133 (*E. coli* and *K. pneumoniae*), PRJ-282721, PRJEB2221, PRJEB5162 (*M. tuberculosis*), PRJEB5225 and PRJEB5261(*S. aureus*). Individual accession numbers and antimicrobial resistance phenotypes are detailed in **Supplementary Data 1**. The distributions of biallelic SNP frequencies are provided in **Supplementary Table 4**.

**Defining the pan-genome.** In order to investigate gene presence or absence we created a pan-genome for each set of isolates. To obtain whole genome assemblies, reads were *de novo* assembled using Velvet[68]. We annotated open reading frames on the *de novo* assemblies for each isolate. We then used the Bayesian gene-finding program Prodigal[69] to identify a set of protein sequences for each *de novo* assembly. These annotated protein sequences were clustered using CD-hit[70] with a clustering threshold of 70% identity across 70% of the longer sequence. We converted the output of CD-hit into a matrix of binary genotypes denoting presence or absence of each gene cluster in each genome (**Supplementary Fig. 2**).

**Kmer counting.** Some diversity such as indels and repeats is difficult to capture using standard variant calling tools. To capture non-SNP variation, we pursued a kmer or word-based approach[18] in which all unique 31 base haplotypes were counted from the sequencing reads using dsk[71] following adaptor trimming and removal of duplicates and low quality reads using Trimmomatic[72]. If a kmer was counted five or more times in an isolate, then it was counted as present, and if not it was treated as absent (**Supplementary Fig. 2**). This produced a deduplicated set of variably present kmers across the dataset, with the presence or absence of each determined per isolate. The total number of SNPs, kmers and gene clusters per species can be found in **Supplementary Table 5**.

**Phylogenetic inference.** Maximum likelihood phylogenies were estimated for visualization and SNP imputation purposes using RAxML version 7.7.6 [73], with a GTR model and no rate heterogeneity, using alignments from the mapped data based on biallelic sites, with non-biallelic sites being set to the reference.

**SNP imputation.** Since Illumina sequencing is inherently more error-prone than Sanger sequencing, strict filtering is required for reliable mapping-based SNP calling, contributing to a small but appreciable frequency of uncalled bases in the genome due to ambiguity or deletion. Restricting analysis to sites called in all genomes is undesirable, while ignoring uncalled sites by removing individuals with missing data at individual sites generates *p*-values that cannot be validly compared between sites because they are calculated using data from differing sets of isolates.

SNP imputation is therefore generally considered necessary for GWAS[60]. We imputed missing base calls using two approaches, ClonalFrameML[53] and Beagle[60]. Imputation using ClonalFrameML[53] involves jointly reconstructing ancestral states and missing base calls by maximum likelihood utilizing the phylogeny reconstructed earlier[61]. To use Beagle the mapped data was coded as haploid (one column per individual), and input as phased data[60,74].

**Testing imputation accuracy.** To simulate data for testing imputation accuracy, 100 sequences were randomly sampled from each GWAS dataset across the phylogeny. Maximum likelihood phylogenies were estimated for the 100 sequences of each species using RAxML[73] as above. Any columns in the alignment corresponding to ambiguous bases in the reference genome were excluded. One round of imputation was performed using ClonalFrameML to produce complete datasets with no ambiguous bases (Ns) that were then treated as the truth for the purpose of testing. The empirical distributions of Ns per site in the datasets of 100 sequences were determined, and these were sampled with replacement to reintroduce Ns to the variable sites in 100 simulated datasets. These sequences were then imputed again using ClonalFrameML and Beagle. Accuracy was summarized per site as a function of the frequency of Ns per site and the minor allele frequency. Overall ClonalFrameML was more accurate than Beagle, thus ClonalFrameML was used for all GWAS analyses (**Supplementary Table 1**).

**Calculating association statistics before controlling population structure.** We wished to compare the significance of associations before and after controlling for population structure. For the SNP and gene presence or absence data, an association between each SNP or gene and the phenotype was tested by logistic regression implemented in R. For the kmer analyses, an association between the presence or absence of each kmer was tested using a $\chi^2$ test implemented in C++. For each variant a *p*-value was computed.

**Correction for multiple testing.** Multiple testing was accounted for by applying a Bonferroni correction[75]; the individual locus effect of a variant (SNP, gene or kmer) was considered significant if its *p*-value was smaller than $\alpha/n_p$ where we took $\alpha = 0.05$ to be the genome-wide false positive rate and $n_p$ to be the number of SNPs and genes, or kmers, with unique phylogenetic patterns, i.e. unique partitions of individuals according to allele membership. Since the phenotypic contribution of multiple variants with identical phylogenetic patterns cannot be disentangled statistically, we found that pooling such variants improved power by demanding a less conservative Bonferroni correction than correcting for the total number of variants (**Supplementary Fig. 10**).

The genome-wide -$\log_{10}$ *p*-value threshold for SNPs and genes (or kmers) was 6.1 (7.3) for *S. aureus* ciprofloxacin, erythromycin, fusidic acid, gentamicin, penicillin, methicillin, tetracycline and rifampicin, 5.9 (6.7) for *S. aureus* trimethoprim, 6.5 (7.3) for all antimicrobials tested for *E. coli*, 6.6 (7.3) for all antimicrobials tested for *K. pneumoniae*, and 5.0 (7.6) for all antimicrobials tested for *M. tuberculosis*. We also accounted for multiple testing of lineage effects by applying a Bonferroni correction for the number of PCs, which equals the sample size *n*.

**Running GEMMA.** For the analyses of SNPs, genes and kmers, we computed the relatedness matrix **K** from biallelic SNPs only. We tested for foreground effects at all biallelic, triallelic and tetrallelic SNPs, genes and kmers. GEMMA was run using a minor allele frequency of 0 to include all SNPs. GEMMA was modified to output the ML log-likelihood under the null and alternative, and -$\log_{10}$ p-values were calculated using R.

To perform LMM on tri- and tetra-allelic SNPs, each SNP was encoded as $K-1$ binary columns corresponding to the first $K-1$ alleles. For each column, an individual was encoded 1 if it contained that allele and 0 otherwise. The first column was input as the genotype, and the others as covariates into GEMMA. The log likelihood of the null from the biallelic SNPs along with the log likelihood under the alternative for each of the SNPs were used to calculate the p-value per SNP.

Due to the large number of kmers present within each dataset, it was not feasible to run LMM on all kmers. Thus we applied the LMM to the top 200,000 most significant kmers from the logistic regression, plus 200,000 randomly selected kmers of those remaining. The randomly selected kmers were used to indicate whether some were becoming relatively more significant than the top 200,000, providing a warning in case large numbers of kmers became significant only after controlling for population structure.

**Variant annotation.** SNPs were annotated in R using the reference fasta and genbank files to determine SNP type (synonymous, non-synonymous, nonsense, read-through, intergenic), the codon and codon position, reference and non-reference amino acid, gene name and gene product.

Unlike the SNP approach where we can easily refer to the reference genome to find what gene the SNP is in and the effect that it may have, annotation of the kmers is more difficult. We used BLAST[76] to identify the kmers in databases of annotated sequences. Each kmer was first annotated against a BLAST database created of all refseq genomes of the relevant genus on NCBI. This enabled automatic annotation of all kmers that gave a sufficiently small e-value against the genus specific database. All kmers were also searched against the whole nucleotide NCBI database, firstly to compare and confirm the matches made against the first database, and secondly to annotate the kmers that did not match to anything in the within-genus database. Finally, when the resistance determining mechanism was a SNP, the top 10000 kmers were mapped to a relevant reference genome using Bowtie2[77]. This was used to determine whether the most significant kmers covered the position of the resistance causing SNP or whether they were found elsewhere in the gene.

Genes were annotated for each CD-hit gene cluster by performing BLAST[76] searches of each cluster sequence against a database of curated protein sequences downloaded from UNIPROT[78].

**Testing power by simulating phenotypes.** To assess the performance of the method for controlling population structure, we performed 100 simulations per species. In each simulation, a biallelic SNP was randomly chosen among those SNPs with minor allele frequency above 20% to be the causal SNP. Binary phenotypes (case or control) were then simulated for each genome with case probabilities of 0.25 and 0.5 respectively in individuals with the common and rare allele at the causal SNP (an odds ratio of 3). For each simulated dataset, we tested for locus effects at every biallelic SNP and lineage effects at every PC, as described above. Power to detect locus effects was defined as the proportion of simulations in which the causal SNP was found to have a significant locus effect. This was compared to a theoretically optimum power computed as the proportion of simulations in which the causal SNP was found to have a significant locus effect when population structure and multiple testing were not controlled for. Power to detect lineage effects was computed as the proportion of simulations in which the PC most strongly correlated to the causal SNP was found to have a significant lineage effect. We defined fine mapping precision as the distance spanned by SNPs within two log-likelihoods of the most significant SNP in the test for locus effects, in those simulations in which the causal locus was genome-wide significant. We calculated the number of homoplasies per SNP by counting the number of branches in the phylogeny affected by a substitution based on the ClonalFrameML ancestral state reconstruction, and subtracting the minimum number of substitutions ($K–1$).

**Software.** We have created an R package *bugwas* implementing our method for controlling population structure and an end-to-end GWAS pipeline using R, Python and C++. Both can be downloaded from [www.danielwilson.me.uk/virulogenomics.html](www.danielwilson.me.uk/virulogenomics.html).

## Acknowledgements
We would like to thank Jean-Baptiste Veyrieras, Deborah Charlesworth and Brian Charlesworth for comments on the manuscript, Xiang Zhou and Matthew Stephens for helping adapt their software, Stefan Niemann for assisting with tuberculosis isolates and Xavier Didelot, Daniel Falush, Rory Bowden, Simon Myers, Jonathan Marchini, Joe Pickrell, Peter Visscher, Alkes Price and Peter Donnelly for useful discussions. This study was supported by the Oxford NIHR Biomedical Research Centre and the UKCRC Modernising Medical Microbiology Consortium, the latter funded under the UKCRC Translational Infection Research Initiative supported by the Medical Research Council, the Biotechnology and Biological Sciences Research Council and the National Institute for Health Research on behalf of the UK Department of Health (Grant G0800778) and the Wellcome Trust (Grant 087646/Z/08/Z). D.W.C. is an NIHR Senior Investigator. D.J.W. is a Sir Henry Dale Fellow, jointly funded by the Wellcome Trust and the Royal Society (Grant 101237/Z/13/Z).

## Author contributions
S.G.E, C.-H.W., J.C., D.J.W. designed the study, developed the methods, performed the analysis, interpreted the results, wrote the manuscript. Z.I., D.A.C. assisted the analysis, commented on the manuscript. N.S., N.C.G., T.M.W., K.L.H., N.W., E.G.S., N.I., M.J.L., T.E.P., D.W.C. designed and implemented isolate collection, drug susceptibility testing and whole-genome sequencing and assisted interpretation. C.C.A.S., G.M., A.S.W. assisted methods development and writing the manuscript.

## Figure Legends
**Figure 1.** Controlling for population structure in bacterial GWAS for fusidic acid resistance in *S. aureus*. (**a**) Effect of controlling for population structure using LMM on significance of the presence or absence of 31bp kmers. The 200,000 most-significant kmers prior to control for population structure and a random 200,000 are plotted. Each kmer is colour-coded according to the PC to which it is most strongly correlated, grey if it is not most strongly correlated to one of the 20

most significant PCs. (**b**) PCs correspond to lineages in the clonal genealogy. Branches are colour-coded by one of the 20 most significant PCs to which they are most correlated. Individual genomes are colour-coded with black or grey lines to indicate fusidic acid resistance and susceptibility respectively. The circle passing through the line is colour-coded to indicate the phenotype predicted by the LMM. (**c**) Wald tests of significance of lineage-specific associations. Some PCs, e.g. PC 9, are hashed to indicate that no branch in the clonal genealogy was most strongly correlated with it. Asterisks above the bars, e.g. PC 25, indicate evidence for lineages associated with particular genomic regions. (**d**) Manhattan plot showing significance ($-\log_{10}$ $p$-values) of unique variants after controlling for population structure, with variants clustered by PC. The horizontal ordering is randomised. This allows identification of the variants corresponding to the most significant lineage-specific associations.

**Figure 2.** Power, false positives, fine mapping and homoplasy in *S. aureus*. Simulation results. (**a**) Controlling for population structure and multiple testing lead to a drastic reduction in power to detect locus effects, compared to the theoretical optimum power for a single locus. The Wald test improves power several-fold by detecting lineage-specific effects. (**b**) Top: Mean number of false positive SNPs and patterns is drastically reduced by controlling population structure with LMM. Bottom: Fine mapping precision is very coarse owing to genome-wide LD. Interpreting lineage effects is useful when the locus-specific signal cannot be fine-mapped. (**c**) The number of times common SNPs (MAF>20%) and antibiotic resistance phenotypes have emerged on the phylogeny. (**d**) When homoplasy is high, power to detect locus effects is much improved, explaining the good power to map antibiotic resistance phenotypes. In the simulations, causal loci were selected at random from high frequency SNPs (MAF>20%) in the $n = 992$ isolates and phenotypes simulated per genome with case probabilities of 0.25 and 0.5 for the common and rare alleles respectively (odds ratio of 3). Genome wide significance (to detect locus effects) was based on a Bonferroni-corrected $p$-value threshold of $\alpha$, equal to 0.05 divided by the number of SNP patterns.

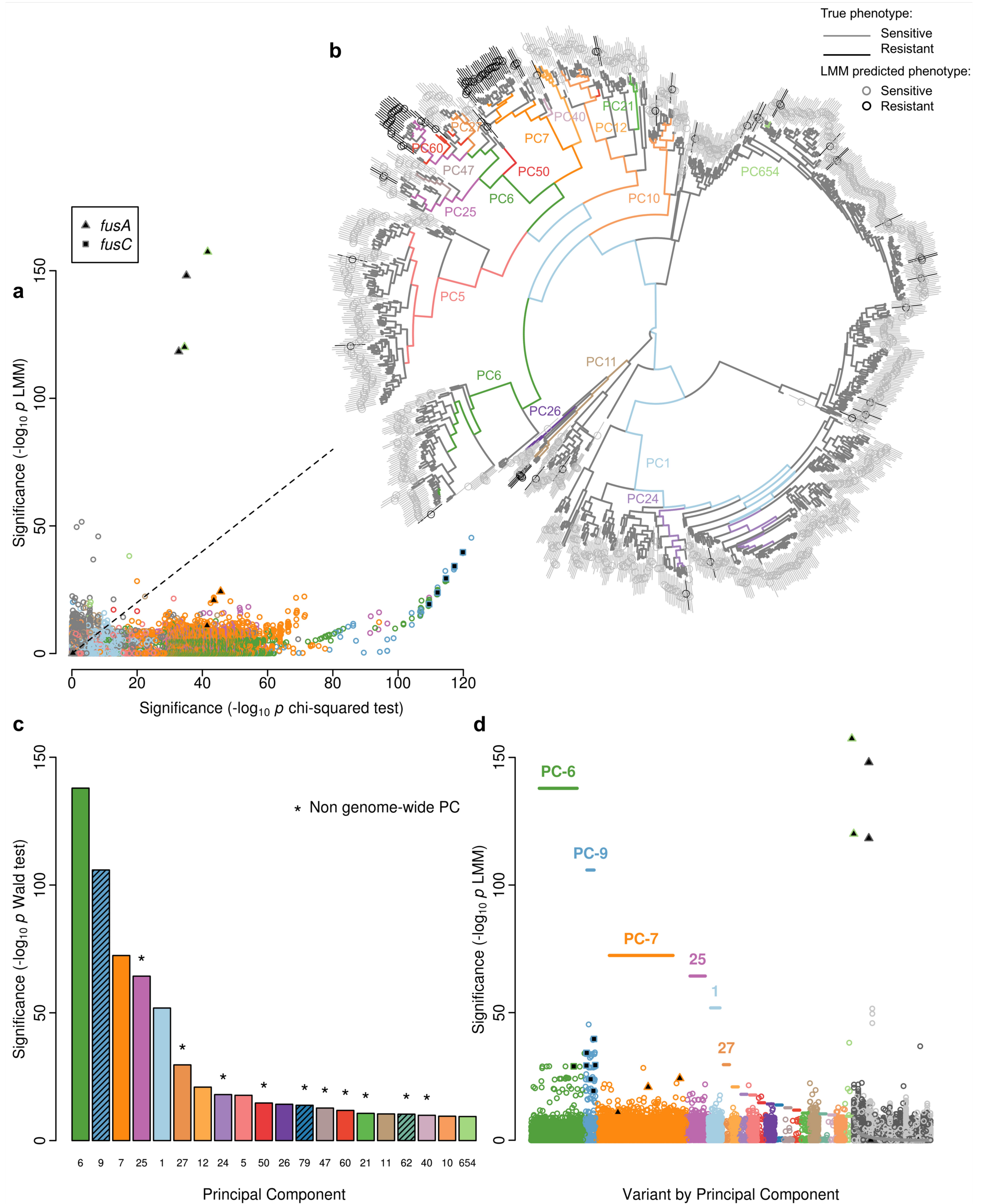

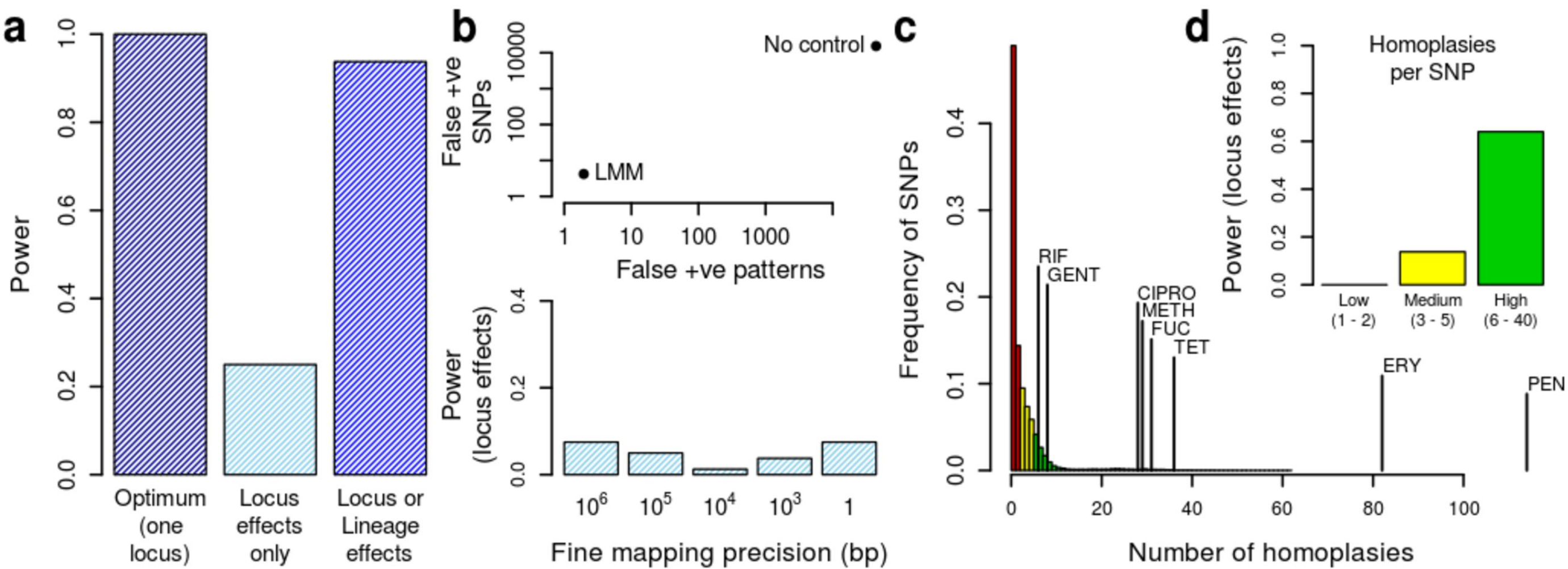

| Antibiotic | # R | # S | Resistance mechanism | SNP / gene rank | SNP / gene LMM rank | Kmer rank | Kmer LMM rank |
|---|---|---|---|---|---|---|---|
| *E. coli* | | | | | | | |
| Ampicillin | 189 | 52 | β-lactamase genes *bla*$_{TEM}$ | 1 | 1 | 6 (tnp) | 6 (tnp) |
| Cefazolin | 139 | 102 | β-lactamase genes *bla*$_{CTX-M}$ | 2 (*nmpC*) | 3 (*nmpC*) | 121710 (nmpC) | 3690 (nmpC) |
| Cefuroxime | 81 | 160 | β-lactamase genes *bla*$_{CTX-M}$ | 1 | 1 | 1598 (162-192 upstream *bla*$_{CMY-2}$) | 470 (162-192 upstream *bla*$_{CMY-2}$) |
| Ceftriaxone | 55 | 186 | β-lactamase genes *bla*$_{CTX-M}$ | 1 | 1 | 1403 (tnp) | 470 (tnp) |
| Ciprofloxacin | 91 | 150 | SNPs in *gyrA*[a], *gyrB*, *parC*[b] or *parE* or presence of PMQR | 1[b] | 1[b] | 1[b] | 1[a] |
| Gentamicin | 48 | 193 | *aac* (*aac(3)-II*), *ant*, *aph* or rRNA methylase | 1 | 1 | 1 | 1 |
| Tobramycin | 67 | 174 | *aac* (*aac(3)-II*), *ant* or rRNA methylase | 1 | 1 | 1 | 1 |
| *K. pneumoniae* | | | | | | | |
| Cefazolin | 53 | 123 | β-lactamase genes *bla*$_{CTX-M}$ | 1 + HP + *wbuC* | 1 | 762 (tnp) | 837 (tnp) |
| Cefuroxime | 46 | 130 | β-lactamase genes *bla*$_{CTX-M}$ | 1 + HP + *wbuC* | 1 + HP + *wbuC* | 762 (tnp) | 1480 (tnp) |
| Ceftriaxone | 35 | 141 | β-lactamase genes *bla*$_{CTX-M}$ | 1 + HP + *wbuC* | 1 + HP + *wbuC* | 771 (tnp) | 812 (tnp) |
| Ciprofloxacin | 34 | 142 | SNPs in *gyrA*, *gyrB*, *parC* or *parE* or presence of PMQR (*qnr-B1*[a], *qnr-B19*[b]) | 2[a] (tnp) | 2[a] (tnp) | 1853[b] (tnp) | 4427[b] (tnp) |
| Gentamicin | 31 | 145 | *aac* (*acc(3)-II*), *ant*, *aph* or rRNA methylase | 1 | 1 | 1 | 79 (*tmrB_2*) |
| Tobramycin | 36 | 140 | *aac* (*acc(3)-II*), *ant* or rRNA methylase | 1 | 1 | 1 | 1 |
| *M. tuberculosis* | | | | | | | |
| Ethambutol | 41 | 1589 | *embB* | 2 (*rpoB*) | 1 | 1 | 1 |
| Isoniazid | 239 | 1470 | *katG*, *fabG1* | 1 | 1 | 1 | 1 |
| Pyrazinamide | 45 | 1662 | *pncA* | 142 (*rpoB*) | 1 | 126 (*rpoB*) | 1 |
| Rifampin | 86 | 1487 | *rpoB* | 1 | 1 | 1 | 1 |
| *S. aureus* | | | | | | | |
| Ciprofloxacin | 242 | 750 | *grlA* or *gyrA* | 1 | 1 | 1 | 1 |
| Erythromycin | 216 | 776 | *ermA*, *ermC*, *ermT* or *msrA* | 1 | 1 | 1 | 1 |
| Fusidic acid | 84 | 908 | SNPs in *fusA*[a] or presence of *fusB* or *fusC*[b] | 4[b] (*SAS0037*) | 1[a] | 75[b] (*SAS0040*) | 1[a] |
| Gentamicin | 11 | 981 | *aacA/aphD* | 1 + GNAT acetyltransferase | 1 + GNAT acetyltransferase | 1 + 415 bases upstream to 100 bases downstream | 1 + 415 bases upstream to 100 bases downstream |
| Penicillin | 824 | 168 | *blaZ* | 1 | 1 | 2 (*blaI*) | 2 (*blaI*) |
| Methicillin | 216 | 776 | *mecA* | 1 | 1 + *mecR1* | 1 + SCC*mec* genes | 1 + SCC*mec* genes |
| Tetracycline | 46 | 946 | *tetK*, *tetL* or *tetM* | 2 (*repC*) | 2 (*repC*) | 1 + plasmid genes | 1 + plasmid genes |
| Trimethoprim | 15 | 308 | SNPs in *dfrB*, presence of *dfrG* or *dfrA* | 1 | 1 | 1 | 1 |
| Rifampicin | 8 | 984 | *rpoB* | 1 | 1 | 1 | 1 |

**Table 1** Number of resistant and sensitive isolates by species and antibiotics, known mechanisms of resistance and main results. The rank of the most significant result for an expected causal mechanism for each GWAS is reported, plus in brackets the gene that was most significant when it was not causal. Where more than one gene or mechanism causes resistance, the variant we found was underlined, or referred to by [a] and [b]. R = Resistant. S = Sensitive. HP = Hypothetical Protein, tnp = transposase. PMQR = Plasmid Mediated Quinoline Resistance. See Supplementary Tables 3-6 for more detail.

- ▭ Resistance determined by gene presence
- ▭ Resistance determined by SNPs
- ▭ Resistance determined by gene presence or SNPs or both
- ▭ Most significant variant was the expected mechanism
- ▭ Most significant variant was in physical linkage (PL) with the expected Mechanism
- ▭ Most significant variant was not the expected mechanism or in PL with the expected mechanism